\documentclass[10pt,conference,letterpaper,twocolumn,twoside]{IEEEtran}
\usepackage[utf8]{inputenc}
\usepackage{amsmath}
\usepackage{amsfonts}
\usepackage{amssymb}
\usepackage{graphics}
\usepackage{graphicx}   
\usepackage{xcolor}
\usepackage{comment}

\usepackage[absolute]{textpos}

\TPMargin{5pt}
\newcommand{\copyrightstatement}{
   \begin{textblock}{0.8}(0.10,0.95)    
        \noindent
        \footnotesize
        Di Li, Asya Mitseva ``Mobile Adhoc Offloading'', in E. Baccelli, F.
        Juraschek, O. Hahm, T. C. Schmidt, H. Will, M.~W\"ahlisch (Eds.),
        Proc.~of~3rd~MANIAC Challenge, Berlin, Germany, July 27 - 28, 2013,
        arXiv, Jan. 2014.
   \end{textblock}
}

\begin{document}

\title{Mobile Adhoc Offloading}

\author{Di Li, Asya Mitseva \\ li@umic.rwth-aachen.de \\Comsys group, RWTH-Aachen University}
\maketitle

\begin{abstract}
This problem is a series of biddings and auctions. Each round of bidding and auction are different from previous ones because of the change of network topology, variance of budget set by the sender, and possible evolution of strategies of other nodes. The huge strategy space of relay nodes makes the formulation to a game very difficult. We present a brief qualitative analysis in this paper, and propose a bidding strategy based on learning algorithms.

\end{abstract}

\copyrightstatement

\section{Preliminaries and Notations}
In this paper, Section I to III describe proposed scheme which competed in MANIAC competition, section IV briefly reports the performance of this scheme, lessons learnt from the competition is concluded in section V.

Both AP and Handhelds auction off the traffic load forwarding service with the same elements: \textbf{budget}, \textbf{fine}, \textbf{time out}, \textbf{source and destination}. 
As the forwarding task has no information of the auctioneer in it, thus downstream entity \footnote{we use entity or node to denote either AP or handheld} adapts strategies without care the auctioneer being AP or handheld. For convenience of analysis, we generalize a recursive auction process as shown in following figure,
	\begin{figure}[ht]
		\centering
	    \includegraphics[width=0.5\textwidth]{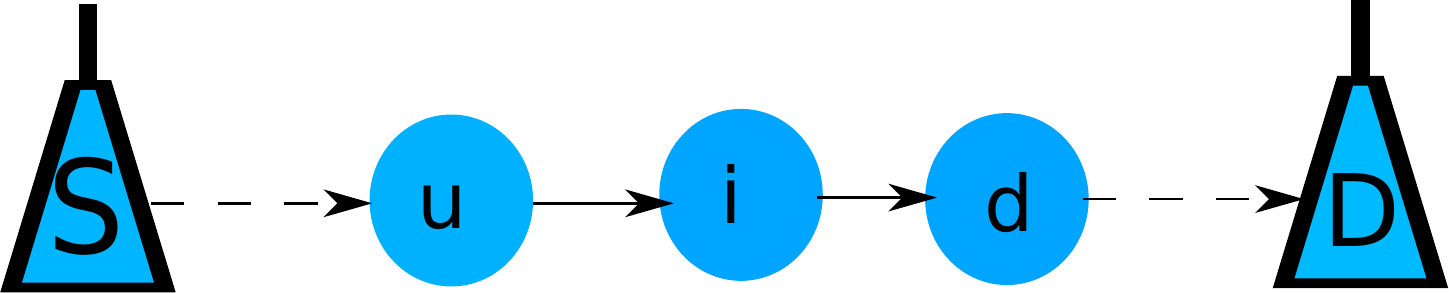}
		\label{recursiveAuction} 
		\caption[]{A recursive auction chain}
	\end{figure}

$S$ and $D$ are APs, which denote source and destination respectively. $u$, $i$, and $d$ represent upstream node, the node being discussed, and the downstream node respectively. Figure 1 shows a generalized situation. We denote the bid to upstream node, budget and fine of advertised auction of a node $i$ with $b_i$, $B_i$, and $f_i$ respectively. 

\section{General Analysis}
\subsection{What should an industrious node do to accumulate money and successful transmissions}

We now analysis the behavior of node $i$. According to rules. when there is a routing request heart, node $i$ has to bid. Clearly, there are two possible outputs for $i$, to win the bid, or to loss. We use a function $P_{\text{winBid}}$ to represent the result of auction. $P_{\text{winBid}}$ has output of 1 and 0, which denote $i$ gets and losses the bid respectively. $P_{\text{winBid}}$ is influenced by the $i$'s bid, along with the bids from the other neighbors of the auctioneer $u$, so $P_{\text{winBid}}$ can be written as $P_{\text{winBid}}(b_i, b_{-i})$, where $b_{-i}$ means the biding strategies taken by the neighbors of $u$ except for $i$. Because of mobility, the set $\{-i\}$ is different in each round.

When node $i$ wins the bid, it starts to consider how to forward the packet. 
We use $P_{\text{taskSucceed}}$ to denote whether the packet is successfully transmitted or not, which is denoted by 1 and 0 respectively.


Now we can express the balance of node $i$ as following\footnote{New rule from competition organizer: An upstream node pays the accepted price to the chosen downstream 
node if the packet has been delivered successfully to the final destination, not immediately after the deal.}
\begin{equation}
\begin{split}
u = \\P_{\text{winBid}}[P_{\text{taskSucceed}}(b_i-b_d) - (1-P_{\text{taskSucceed}})(f_u-f_i)] = 					\\
																					\\
\left\{ \begin{array}{ll}
f_i - f_u 	& \mbox{if $P_{\text{winBid}}=1, P_{\text{taskSucceed}}=0 $}	\\
b_i - b_d 				& \mbox{if $P_{\text{winBid}}=1, P_{\text{taskSucceed}}=1 $}	\\
0						& \mbox{if $P_{\text{winBid}}=0$}							\\
b_i - B_S				& \mbox{if $P_{\text{winBid}}=1$, then piggyback}										\\
b_i - f_u				& \mbox{if $P_{\text{winBid}}=1$,  then do nothing}
\end{array}
\right.
\end{split}
\end{equation}


From the above possible outcomes of different situations, we can see that the first and last scenarios produce positive balance when $f_u$ is not very high, but there is little possibility to be allowed by competition organizers. The second scenario where $i$ wins bid and afterwards successfully transmit the packet causes profitable outcome for $i$, thus becomes the most favorable situation. In order to achieve this, node $i$ needs good strategy to struggle for the bid, and then wisely choose the next hop which helps transmit the packet successfully with a higher possibility.

\section{Sketch of Strategy}

\subsection{Auction}
Find next hop to forward packet. Neighbors' ability of forwarding packet is decided solely by time out time $timeout$ and the distance between neighbor and destination $dist$.

Working with OLSR routing scheme, when a node has a packet to forward, it knows the minimum number of hops (denoted as $dist$) to the destination by looking at its routing table (the hop counts in the corresponding entry). It is easy to know that, among its neighbors, the maximal minimum number of hops can not exceed $dist + 1$. 
\begin{itemize}
\item If $timeout \geq dist + 1$, delivery is an easy task, choose the neighbor with the smallest bid.
\item If $timeout = dist$, delivery is a risky job, choose the next hop which is closet to destination in order to improve the possibility of successful delivery.
\item If $timeout < dist$, delivery is mission impossible, set budget randomly, choose any neighbor, and set fine as high as budget.
\end{itemize}

The budget is set as follows,
\begin{equation}
\label{budgetSetting}
\begin{split}
\left\{ \begin{array}{ll}
b_i\times dist/timeout & \mbox{if $timeout \geq dist$} \\
b_i & \mbox{if $timeout < dist$} 
\end{array}
\right.
\end{split}
\end{equation}
where $b_i$ is the biding price of $i$ to win the bid from upstream node. The idea is if timeout is larger than $dist$, which means this forwarding task can be finished safely and the motivation of forwarder is thus higher, then nodes will still be happy to forward the packet even with less payoff. 
\subsubsection{Fine}
If budget is too low (the fine is accordingly low), the next hop may possibly drop the packets and cause huge loss for $i$ whereas its own loss is limited. If we set $f_i$ as $b_i-\epsilon$, where $\epsilon$ is a small value, then the budget should be $\max\{f_u/2, b_i\times dist/timeout\}$, in this case, if next hop node $d$ drops the packet, it will pay at least $f_u/2$ back to you, and thus both of $i$ and $d$ loss the same amount (You avoid being played by malicious node).

\subsection{Bid}
For node $i$, if the $timeout$ in the routing request from node $u$ is smaller than its $dist$, then $i$'s bidding price will be set as $B_u$. The idea behind is that $i$ is not willing to be chosen for this \textit{mission impossible}. If $timeout \leq dist$, $i$ needs to make effort to win the bid. The only way to win is to bid with smallest bidding price, to achieve this, we combine two learning schemes.

\subsubsection{Prediction with supervised learning and historic records}
In this scheme $i$ needs to predicts the bidding prices of the other nodes hearing the same request (its competitors in this auction). This task is hard to do as some competitors may not be $i$'s neighbors and thus $i$ is unaware of them, but we argue that $i$'s neighborhood provides adequate similar samples to predict auctioneer's neighbors after winning \textit{several} bid successfully. In order to do so, $i$ needs to collect the pair of routing request from auctioneer and corresponding bidding prices. We assume that there exit a fixed pattern that the other nodes deciding their bidding prices based on the heart budget. Exactly speaking, $i$ maintains a table for all the other nodes about their bidding history. Then a simple machine learning scheme (regression on the basis of training examples obtained in previous auctions) is used here to obtain the possibly minimum bidding price $p_{-i}^{min}$ of others. Then $i$ will choose its bidding price $b_i$ slightly smaller than $p_{-i}^{min}$. If $i$ fails to win this bid, it will become more aggressive (with lower bidding price) to compete next time, which means the price will be a certain factor higher.


There exists drawback in this scheme, which is in the initial phase of prediction, there is only a few samples can be used for regression, thus we propose the other scheme to improve.

\subsubsection{Regret learning scheme}

There are several components in this unsupervised learning scheme:
\begin{itemize}
\item We set 10 biding price levels $b_{i,1}, b_{i,2}, b_{i,3}, b_{i,4}.., b_{i,10}$, where $b_{i,x} = B*x/10$, in order to work with finite solution space.
\item $\mathcal{R}$: Regret matrix, $10\times 10 $, which stores regret value from different actions.
\item $Potential$: Potential set, $10\times 10 $, $Potential(a,b)$ is the change of utility when $i$ changes price from action $a$ to $b$.
\item Regret matrix evolves with the game continues:
\[\mathcal{R}^{r+1}_i = (1-\frac{1}{r+1})\times \mathcal{R}^{r}_i + Potential^r\]
\end{itemize}

where $r$ is the packet forwarding round index.

$\mathcal{R}^{1}_i$ is randomly decided, and then involves based on the above formula. We design a heuristic way to decide $Potential$: 

Assume in the first round ($r =1$), node $i$ chose bid price $b_{i,3}$ and fails to win, then it won't help if $i$ increased its biding price, so we set potential of this change as negative $x-3$ where $x <3$, if $i$ increase its bidding price, then it will possible for $i$ to win, so the potential is set as $Potential(3,x) = 10-x$ where $x >3$.

When $i$ wins the bid with bidding price $b_{i,3}$, it can secure the bid by decreasing its biding price (using $b_{i,1}, b_{i,2}$), but its income will be decreased, thus decreasing its biding price is not favored, we set $Potential(3,x)= x-3$ where $x < 3$, when $i$ increases its biding price, it is possible to lose the bid, so potential is also set to be negative as $Potential(3,x)= 3*(3-x)$ when $x > 3$. Note we add a coefficient 3 here to emphasis the serious outcome of increasing biding price.


 In each round $r$, node $i$ choose the biding price which poses the biggest regret in the regret matrix.

\subsubsection{Combine the results from two schemes together}

The adopted bidding price of $i$ is  $\min\{b_{i,x},B_u\times (dist/timeouts)^n\}$.

\subsection{Behave aggressively}

The above analysis doesn't consider the ultimate purpose of the competition: to defeat all the other competitors with more money earned (or packets successfully delivered), instead,  it only focuses on its own welfare. To win the competition, node $i$ should act more selfishly to avoid helping greedy competitors (upstream node which set low budget) make big money whereas itself only gets the changes. So in case the budget is low, $i$ would simply drop the packet and cause big loss to the upstream (may not be the immediate previous one) and endure a slight loss in the same time.

\subsubsection{Drop packet when the budget is too low}
\label{aggressive_drop}
Question: If I drop the packet purposely, will I get revenged maliciously some time later?

as the the price of revenge is not trivial (the node taking revenge may need to pay considerable fine upwards), so revenge is not an good option for any node. In a word, if we can deduce there is one greedy upstream node exists, we can safely drop the packet.

We assume all the other participates have the same conclusions as us.

\subsubsection{Avoid giving bid to 'rich' node}
When being auctioneer, we try to not give the bid to the component which is \textit{successful}. To achieve this, we need to keep records of each node which wins a bid from the very beginning of the competition. Based on the same argument for biding price prediction, we assume any other node winning a bid is heart by either of our two devices.

For our node $i$, it maintains tables of revenues for all rivals. For example, the table for rival $a$ on node $i$ is:

\begin{tabular}{|c|c|c|}
\hline  transactionID & auctioneerID & revenue \\ 
\hline  2 & m & $B_m*\mu-f_m*(1-\mu)$ \\ 
\hline  $\cdots$ & $\cdots$ & $\cdots$ \\ 
\hline  7 & n & $B_n*\mu-f_n*(1-\mu)$ \\ 
\hline 
  \end{tabular} 

where $\mu$ is as follows,
\begin{equation}
\begin{split}
\left\{ \begin{array}{ll}
timeout/dist & \mbox{if $timeout < dist$} \\
1 & \mbox{if $timeout > dist$} 
\end{array}
\right.
\end{split}
\end{equation}
Note that according to the rules, in one transaction (packet transmission), any node can at most win bid for once, that is why there is only one auctioneerID in each row. Two devices from the same team merge their table together when they become neighbours.

\subsection{Communication between team mates}
\begin{itemize}
\item share history records on others' bidding prices and accumulated revenues.
\item give team mate priority when deciding who will win the routing req from it.
\end{itemize}

\section{Performance in competition}
\subsection{Introduction of the real competition setting}
The access points are deployed in two floors in one teaching building. The mobile ad hoc network is composed with 10 tablets from 5 participant teams and two tablets from organization team. Each tablet is held by one person who walks randomly and freely in the two floors covered by signals from access points. There are totally 50 access points, and each tablet averagely catches 3 to 8 access points. There are 3 rounds of competitions. Each lasts 10 minutes and there is time between rounds for adjustment.

\subsection{Performance}
In the first round we didn't adopt the aggressive model (\ref{aggressive_drop}) and exclude the possibility that the packets sent from access points are with small budget, because we thought this would discourage the forwarding willingness of nodes and finally result in low ratio of successful transmission. Unfortunately, large amount of packets with small budget were seen. 
Our tablets won most bids. The new budget set in our routing request is decided by Formula \ref{budgetSetting}, as \textit{timeout} used in the competition is big (20, meanwhile the number of hops is 1 to 4 or 5), thus the new budget is very small even the won packets are with high budget. As a result, most packets forwarded by us are dropped by next hops. In one word, our tablet suffered a big loss of fine. 

In the second and third rounds, we dropped the packets with small budget (smaller than 30\% of the maximal budget) thus experienced minor loss. After winning a auction, we set higher budget and accordingly high fine for the packet, (more than 50\% of the previous budget), which refrains the downstream nodes to drop packets easily. Although deficit is alleviated compared with the first round, we noticed the low success ratio of transition. The reason is we didn't pay attention to the forwarding ability, or forwarding willingness of downstream nodes, so that the transmission failed and we got fined.

\section{Lessons learnt}
\begin{itemize}
\item Auction strategy

$\textbf{For participants}$: In case the budget is low, raise bid price to avoid being chosen, in case the budget is high, make sure the new budget is not too small.

$\textbf{For system}$: packets sent from APs should not be assigned with low budget.

\item Bidding strategy

We won a big fraction of routing quests that are heart, which illustrates our bidding strategy works quite well, and is adaptive to other nodes' bidding behaviours.

\item Choosing next hop

The willingness of next hop node should be considered carefully, our scheme which chooses next hop solely based on bidding prices doesn't perform well. Although the mechanism deciding the willingness is unknown, it is safe to assume the deciding mechanism is static, then it is possible to evaluate nodes' willingness for forwarding based on historical record. Based on the setting of competition, we are aware whether the next hop node forwards packet by waiting for the $\textbf{BID\_WIN}$ sent from the next hop. We give each other node one value called \textbf{willingness}. If $\textbf{BID\_WIN}$ is not heart, which means the packet is either gets dropped, or the next hop doesn't have neighbours, then we label this node as non-cooperative and $\textbf{willingness}-=1$, or cooperative if the $\textbf{BID\_WIN}$ is heart, then $\textbf{willingness}+=1$. This \textbf{willingness} can be used as metric to choose the next hop in next transmission. As the ad hoc network is small and dynamic, it is easy to accumulate considerable number of records for each neighbour. A more complex and possibly efficient way could be: find a function deduced from history, which produces the probability that the node is cooperative or not given a budget, then decision can be made based on the probability in next transmission.


\item Participates only care their own interests, and system-wide consideration is not necessary and improper.

All the related work recently \cite{gameRoutingSurvey2008} discuss forwarding strategies with an assumption that, all the nodes have the same deciding mechanism on auctioning and bidding. As participating teams exploit orthogonal mechanisms and common notions don't exist, it is difficult for ingenious algorithms to achieve good performances in this competition. This competition can be seen as a highly demanding scenario for forwarding strategies.

\end{itemize}

\bibliographystyle{abbrv}		
\bibliography{../masterarbeit/myrefs}
\end{document}